\documentclass[twocolumn,aps,prl,nobibnotes,byrevtex,nofootinbib,superscriptaddress,amsfonts,amsmath,amssymb,floatfix,10pts]{revtex4-1}
\usepackage{appendix}
\usepackage{hyperref}
\usepackage{enumerate}
\hypersetup{
    colorlinks=true,
    citecolor=blue,
    linkcolor=blue,
    filecolor=magenta,      
    urlcolor=blue,
}

\usepackage{charter}
\usepackage{subfig}
\usepackage{graphicx}
\usepackage{multirow}
\usepackage{dcolumn}
\usepackage{mathrsfs}
\usepackage{acronym}
\usepackage[usenames,dvipsnames,svgnames]{xcolor}
\usepackage{soul}
\usepackage{tabularx}

\begin{document}
\title{Low-Frequency Noise Mitigation and Bandgap Engineering using Seismic Metamaterials for Terrestrial Gravitational Wave Observatories}
\author{John J. Oh}
\email[]{johnoh@nims.re.kr}
\affiliation{Gravity Research and Application Team (GReAT), Division of Fundamental Research on Public Agenda, National Institute for Mathematical Sciences, Daejeon 34047, Republic of Korea}

\date{\today}

\begin{abstract}
Gravitational-wave now became one of the important observational methods for studying the Universe since its first detection. However, the ground-based observatories have an inherent barrier to their detection frequency band due to the seismic and gravity gradient noises nearby the perturbation of the surroundings. A recent intriguing development of artificial structures for media called metamaterial is opening a new branch of wave mechanics and its application in various fields, in particular, suggesting a novel way of mitigating noises by controlling the media structure for propagating waves. In this paper, we propose a novel framework for handling noises in ground-based gravitational wave detectors by using wave mechanics under metamaterial media. Specifically, we suggest an application of the bandgap engineering technique for mitigating the underground effects of acoustic noises resulting from the seismic vibration in the KAGRA gravitational wave observatory.
\end{abstract}

\maketitle

\acrodef{gw}[GW]{gravitational wave}
\acrodef{ligo}[LIGO]{Laser Interferometer Gravitational-wave Observatory}
\acrodef{snr}[SNR]{signal-to-noise ratio}
\acrodef{mm}[MM]{metamaterial}
\acrodef{dps}[DPS]{double positive medium}
\acrodef{eng}[ENG]{$\epsilon$ negative medium}
\acrodef{dng}[DNG]{double negative medium}
\acrodef{mng}[MNG]{$\mu$ negative medium}
\acrodef{fem}[FEM]{finite element method}
\acrodef{em}[EM]{electromagnetic}
\acresetall


\ac{ligo} was the first to detect \ac{gw} produced by merging binary black holes on September 14, 2015 \cite{Abbott:2016blz}. This suggests that we now have a new tool for investigating the Universe by detecting gravitational changes in the history of observing the Universe, which previously depended on optical and \ac{em} approaches, culminating in the creation of a new science known as \ac{gw} astronomy. As a result of this discovery, numerous international efforts to study \ac{gw}s are now accelerating to create novel ideas for future \ac{gw} telescopes, including the ground-based \ac{gw} observatories like the Cosmic Explorer \cite{LIGOScientific:2016wof} and Einstein Telescope \cite{Punturo:2010zz} as well as the space-based \ac{gw} antennas like LISA \cite{Vitale:2014sla}.

While space-based \ac{gw} antennas are made to detect signals such as merging supermassive black holes in the submillimeter Hz frequency band, ground-based \ac{gw} telescopes are primarily made to detect \ac{gw}s from binary collisions in the $30-2000$Hz frequency range. The difference between ground-based \ac{gw} telescopes and space-based \ac{gw} antennas is the presence of a seismic noise barrier brought about by the activity of the Earth. Seismic noise below $30$ Hz resulting from the activities of the ground on Earth appears to be an inherent limitation that ground-based \ac{gw} telescopes encounter.

Since terrestrial \ac{gw} telescopes are installed on the ground, a variety of environmental conditions, including earthquakes, lightning, wind, and temperature, have an effect on their ability to be detected \cite{Aasi:2014mqd, TheLIGOScientific:2016zmo}. Furthermore, it is necessary to include noise effects from gravity gradients such as microseismic variations and acoustic disturbances caused by Earth's activity and continuous change in air pressure, temperature, and other factors \cite{Saulson:1984yg, Hughes:1998pe, Beccaria:1998ap, Beker:2012zz}.
Therefore, overcoming the limitations of terrestrial \ac{gw} telescopes need new paradigms for understanding and reducing the surrounding noise factors affecting them.
Installing the \ac{gw} telescope underground to limit the transmission of surface vibrations, such as the Einstein Telescope and KAGRA \cite{Akutsu:2018axf}, is an effort to avoid some locations of the seismic noise barrier. Because there is significantly less seismic activity and no gravity perturbation resulting from the convection flow by the atmosphere and ocean, another new concept against this limitation is to put an \ac{gw} antenna on the Moon's surface \cite{LGWA:2020mma}.

Apart from the aforementioned efforts, ground-based \ac{gw} telescopes must decrease environmental noise like seismic vibrations to improve data quality and \ac{gw} detection efficiency. Various noise sources impacting the \ac{gw} telescope propagate as waves; for example, a strong transient seismic wave near a \ac{gw} observatory breaks the locked state of the laser interferometer, causing the observation mode to halt. 
In the network observation of \ac{gw}, this interruption of the observation mode is a drawback for both the combined \ac{snr} and the sky localization efficiency of the \ac{gw} event. In this point of view, it is of great importance to understand various noise sources and mitigate them for improving \ac{gw} detection performance. In this study, we address the possibilities of minimizing this ambient noise by modulating wave propagation from the point of view of \ac{mm} science.
\begin{figure}[t!]  
\begin{center}
\includegraphics[width=\columnwidth]{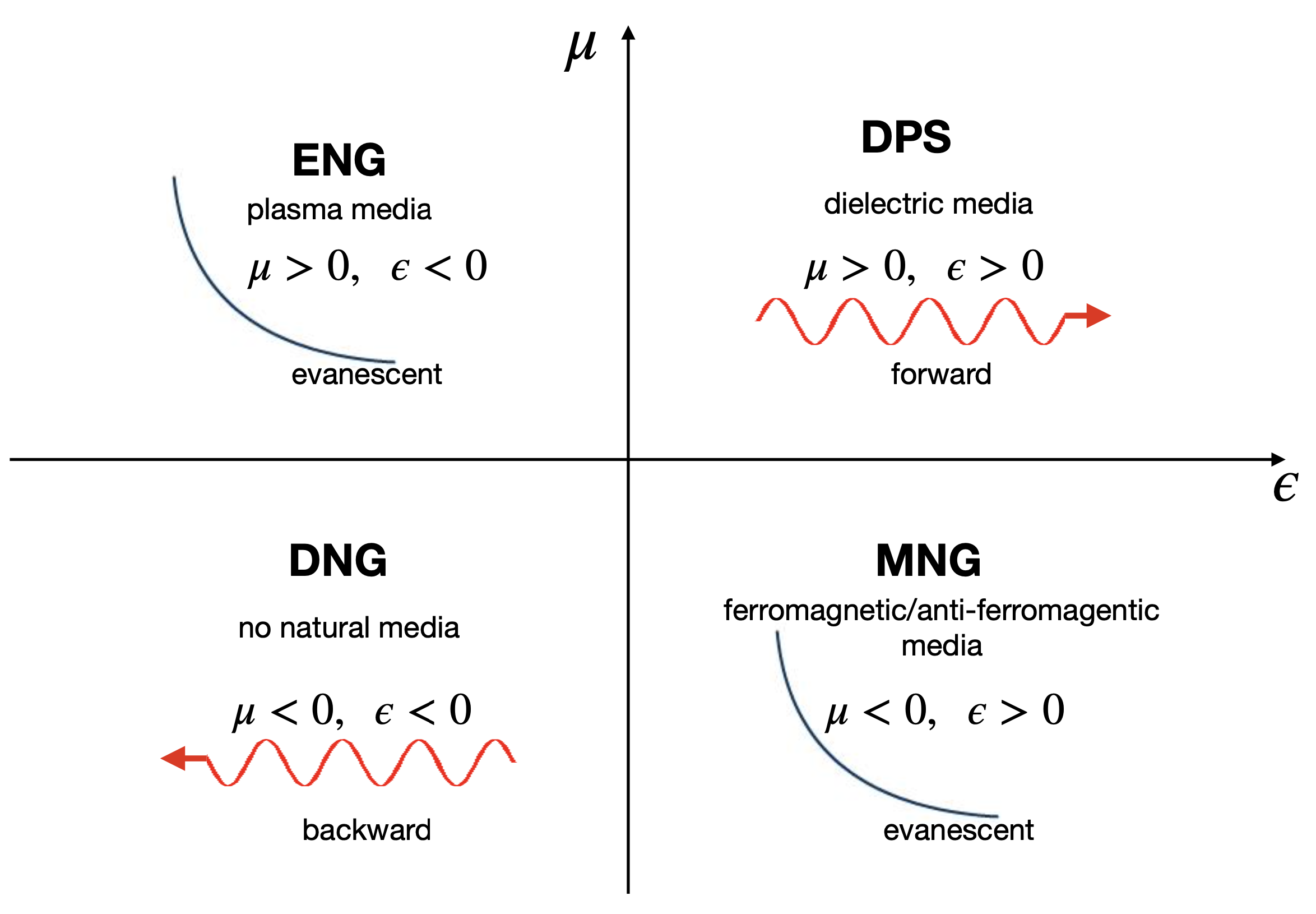}  
\caption{\small A graphical illustration of the properties of the media in the $\epsilon$ and $\mu$ space for the \ac{em} waves. We are familiar with the \ac{dps} that yields the usual feature of \ac{em} wave propagation. The \ac{eng} and \ac{mng} can be realized and found for a specific medium in nature. However, the \ac{dng} state does not exist in nature. The artificial media structure with this state is called {\it metamaterial}.} \label{fig:media}
\end{center}
\end{figure}

\begin{figure*}[ht!]  
\begin{center}
\includegraphics[width=\textwidth]{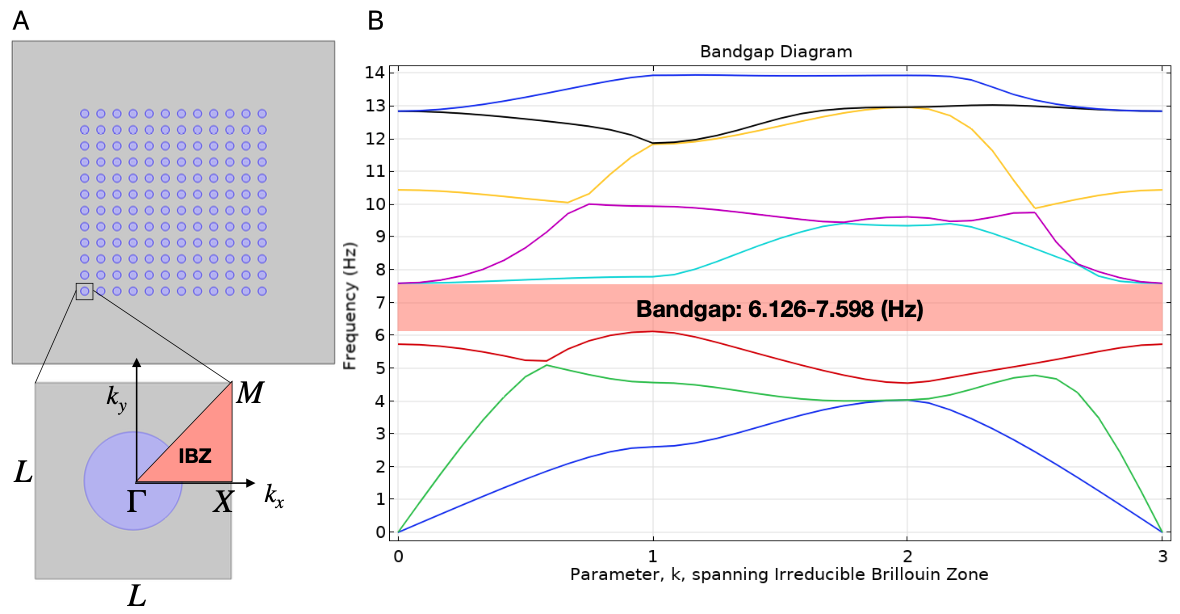}  
\caption{\small A two-dimensional example of periodic structural steels (blue circle) on soil platform (grey domain); (A) Each unit cell describes an irreducible Brillouin zone (IBZ) with a Bloch-Floquent boundary condition. (B) Solving the eigenvalue equation produces a bandgap structure between $6.126-7.598$ Hz in $f-k$ space.} \label{fig:cell}
\end{center}
\end{figure*}

\ac{mm} is defined as an {\it artificial} structure of media that causes unconventional properties of wave propagation. Historically, it was firstly introduced by Veselago \cite{Veselago_1968} that the \ac{em} fields can unconventionally propagate under a specific media condition, possessing a special property such as a negative refraction index, $n=\sqrt{\mu \epsilon}$, by controlling the dielectric constant, $\epsilon$, and the magnetic permeability, $\mu$. Because we have a dispersion relation, $k=\sqrt{\mu\epsilon}\omega$, then it is found that the wave propagates when $\mu\epsilon >0$ whereas it decays exponentially when $\mu\epsilon<0$. Here, we have four possible cases of wave propagation; \ac{dps}, \ac{eng}, \ac{dng}, and \ac{mng} as below:
i) \ac{dps} material ($\mu>0$ and $\epsilon>0$): forward wave propagation of most dielectric media ii) \ac{eng} material ($\mu>0$ and $\epsilon<0$): evanescent waves for plasma media iii) \ac{mng} material ($\mu<0$ and $\epsilon>0$): evanescent waves for gyrotropic magnetic media iv) \ac{dng} material ($\mu<0$ and $\epsilon<0$): backward wave propagation but no such media in nature.
The classification of the medium can be illustrated graphically as shown in Fig. \ref{fig:media}.
Many intensive studies have been conducted for the practical realization of \ac{dng} for \ac{em} waves using an artificial circuit structure \cite{798002, doi:10.1126/science.1058847, doi:10.1126/science.1125907}, leading to the emerging development of metamaterial devices in nanophotonics. The \ac{eng} and \ac{mng} states, which produce an evanescent wave, may be used to reduce wave amplitudes in a variety of fields in addition to the realization of \ac{dng}. Refer to \cite{10.5555/1592927} and references therein for the basic theory and its application to \ac{mm}. 

On the other hand, similar wave properties with \ac{em} \ac{mm} were found to also be valid when controlling the effective density, $\rho$, and the effective compressibility, $\chi$, for acoustic waves \cite{2000Sci...289.1734L, 2000PhRvL..84.4184S,doi:10.1142/9789813228702_0002}, and the mass density, $\rho$, and the bulk modulus, $B$, for seismic waves \cite{2014PhRvL.112m3901B, 2016NatSR...619238C, SRL2018Roux, doi:10.1142/9789813228702_0007}. The comprehension of \ac{mm} properties was understood by investigating the structure of the band gap in photonic crystals \cite{PhysRevLett.58.2059} and photon localization \cite{PhysRevLett.58.2486}. For acoustic and elastic waves, a similar bandgap structure was studied in phononic crystals \cite{PhysRevB.48.13434, PhysRevLett.71.2022}. The theoretical formulation of bandgap structure for \ac{em}, elastic, and acoustic waves has been extensively studied in Refs. \cite{SIGALAS1992377, Pendry1994, PhysRevE.52.1135, Poulton2000}. 
Many practical applications of metamaterials can be considered in acoustics and seismological engineering, such as acoustic insulation \cite{PhysRevLett.71.2022}, a breakwater device for ocean waves \cite{PhysRevLett.95.154501}, and earthquake engineering \cite{2014PhRvL.112m3901B, BRULE2020126034,Miniaci_2016,Colombi:2016:10.1038/srep27717}. 

\begin{figure*}[t!]  
\begin{center}
\includegraphics[width=\textwidth]{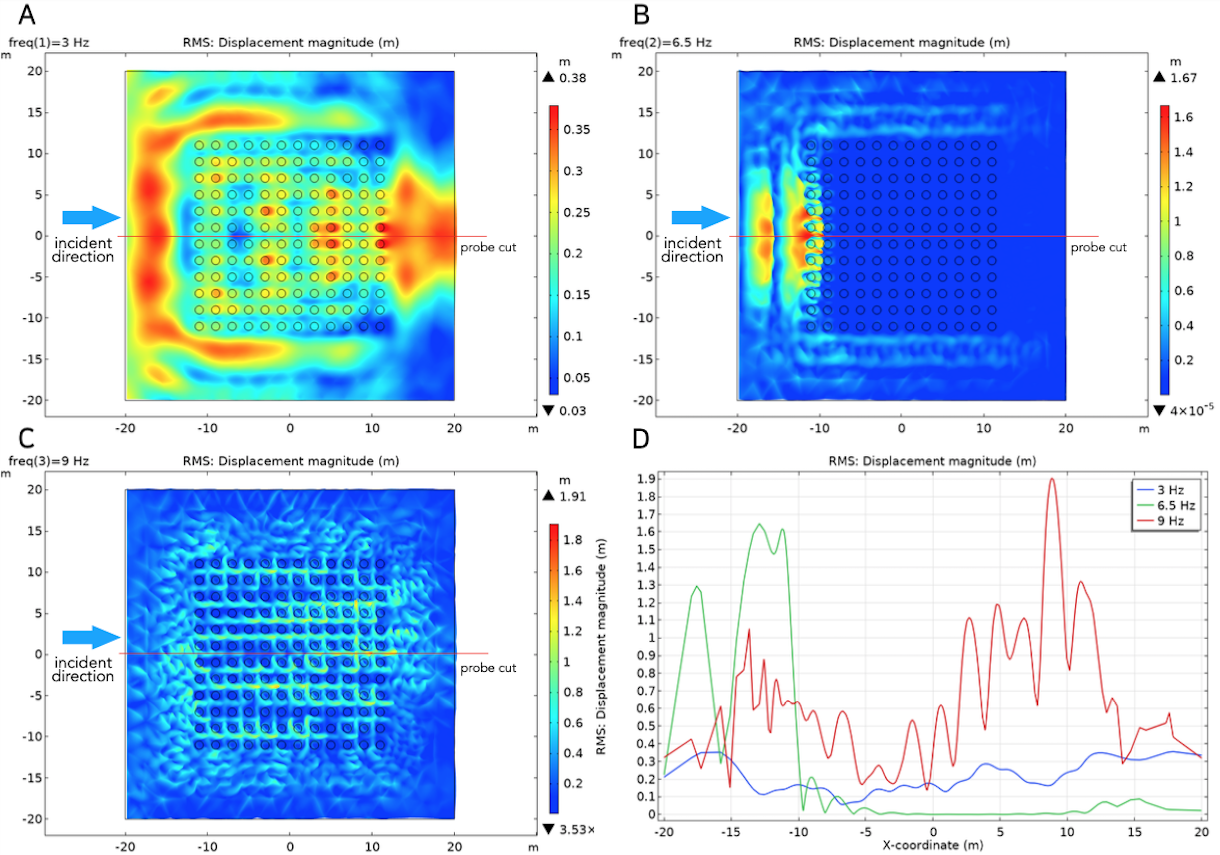}  
\caption{\small The root-mean-squared (RMS) displacement of magnitude by the solid mechanical simulation on the platform with soil and a periodic array of structural steel beams (A, B, C). The incident displacement is applied from the LHS boundary for the $\{3, 6.5, 9\}$ Hz frequencies. The RMS displacement magnitude along the probe cut line at $(x, 0)$ is drawn for the considered frequencies (D), which shows that the displacement amplitude can be reduced by the metamaterial-like structure in the bandgap frequency (the green curve at $(-11, 11)$ [m]).} \label{fig:2Dsimuls}
\end{center}
\end{figure*}

In this point of view, we investigate the feasibility of applying \ac{mm} to reduce the low-frequency noise resulting from seismic and/or acoustic waves that are affected in the ground-based observatory \ac{gw}, in particular, the Japanese \ac{gw} detector called KAGRA. The KAGRA detector is an underground 3 km laser interferometer with a cryogenic test mass configuration. In Ref. \cite{Jung:2022kyc}, it was addressed that the underground effect by wind can be affected by the air pressure perturbation in the KAGRA gravitational wave detector tunnel in the low frequency band. More specifically, a close association between wind speed during the day on the Ikeno mountain has been reported and a nonlinear correlation found between \ac{gw} and microphone channels at the same time. A hypothetical explanation is a transformation from seismic wave propagation to acoustic vibration in the tunnel. As addressed in \cite{Jung:2022kyc}, the noise from the sound pressure inside the tunnel in the directions of the $y$ arm and the $x$ arm is quite different from each other due to the configuration of the KAGRA detector inside the mountain; the $y$ arm is parallel to the slope of the valley, while the $x$ arm is positioned from the slope to deep inside the mountain. The strong wind in the valley of the mountain strikes the slope of the $y$ arm side, then the seismic wave propagates from the slope to the tunnel of the KAGRA detector. Its vibration can transform into an acoustic wave inside the tunnel, which can produce acoustic noise. A simplified finite element analysis has been shown in \cite{Jung:2022kyc}, which presents a plausible explanation for the coincidence during the windy period. Since low-frequency acoustic noise may be thought of as transforming from seismic waves, it is anticipated that decreasing seismic vibration will necessarily result in a reduction in acoustic noise.
In this background, we want to mitigate the seismic propagation by the wind that transforms into acoustic vibration in a low-frequency band through metamaterial structures and band-gap engineering.
\begin{figure*}[t!]  
\begin{center}
\includegraphics[width=\textwidth]{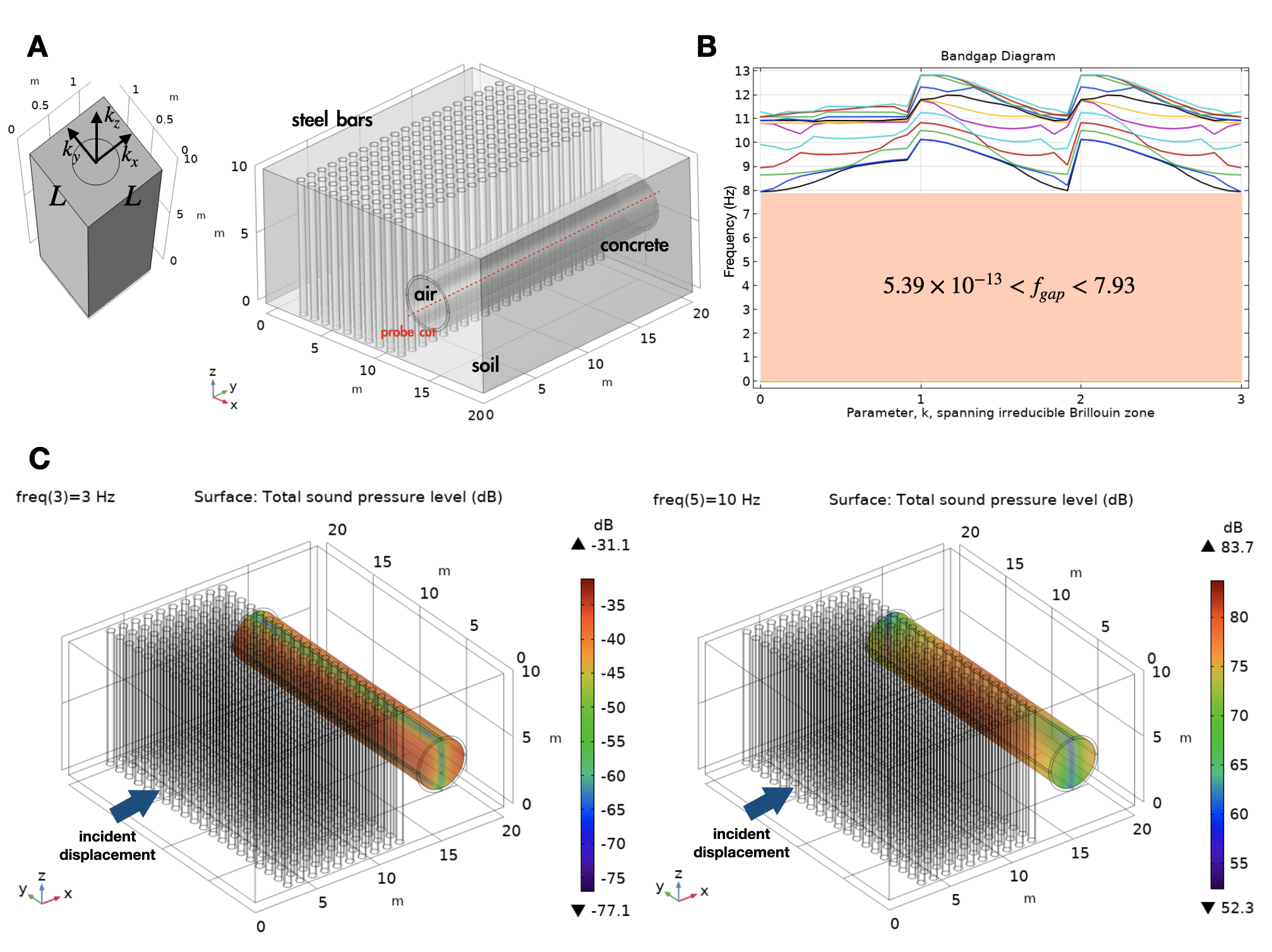}  
\caption{\small (A) The three-dimensional platform of soil, structural steel bars and the underground concrete tunnel, (B) bandgap structure of $5.39\times 10^{-13}~{\rm Hz} < f_{{\rm gap}} < 7.93 ~\rm Hz$, (C) The solid acoustic multiphysics simulation for the sound pressure level inside the tunnel caused by seismic disturbance in the vicinity of the metamaterial-like beam structure} \label{fig:KAGRA-MM1}
\end{center}
\end{figure*}

Let us start with a simple case of a two-dimensional plane with periodic structure as shown in Fig. \ref{fig:cell}A. The specification of each circle is one meter in diameter and is squarely arranged in $11\times 11$ square meters at intervals of one meter each. Each circle is assumed to be structural steel with typical material properties of $\rho={7.85\times 10^{3} ~{\rm[kg/m^3]}}$, $B={2.0\times 10^{11}~{\rm [Pa]}}$, and the Poisson ratio $\nu={0.30}$. The residual domain is assumed to be a soil with the material properties of $\rho={1.35\times 10^{3}~{\rm[kg/m^3]}}$, $B={1.0\times 10^{6}~{\rm [Pa]}}$, and $\nu={0.45}$. A unit cell containing a structural steel in Fig. \ref{fig:cell}A describes an irreducible Brillouin zone (IBZ) with a Bloch-Floquet boundary condition. The boundary constraints the displacement at boundaries of the periodic structure as $u_{2} = e^{-i\vec{k}_{BF}\cdot(\vec{x}_2-\vec{x}_1)} u_{1}$, where $\vec{k}_{BF}$ is the Bloch-Floquet wave vector. We impose the periodic boundary condition on the edge of the unit cell, then the eigenvalue problem on this IBZ cell, 
\begin{equation}
    \nabla^2 u_{z} + \frac{\rho}{\mathfrak m} \omega^2 u_{z} = 0,
\end{equation}
completely determines the dispersion relation and the bandgap structure, where $\mathfrak m$ is a Lam\'e constant. As a result, the bandgap structure between $6.126$ Hz and $7.598$ Hz is shown in Fig. \ref{fig:cell}B. 


On the other hand, we consider the deformation of a solid platform by an external force for a given material, which is described by the continuum equation of motion as
\begin{equation}
    \label{eq:conteq}
    \rho \frac{\partial^2 {\bf u}}{\partial t^2} = {\bf F}_{v} - \nabla_{X}\cdot{\bf P}^{T},
\end{equation}
where ${\bf P}$ is the first Piola-Kirchhoff stress tensor and ${\bf F}$ is the deformation gradient. Here, the divergence can be calculated with respect to the coordinate system in the material frame and ${\bf F}_{v}$ is the volume force vector. Note that Eq. (\ref{eq:conteq}) can be equivalently expressed in terms of the Cauchy stress tensor, ${\bf \sigma}$, with ${\bf P}=J{\bf \sigma}{\bf F}^{-T}$ with respect to the spatial coordinate, where $J$ is the Jacobian of ${\bf F}$, $J=\det{\bf F}$. The relevant Lam\'e constant in this configuration for material properties is given as $\lambda=1.15\times 10^{11}~{\rm [Pa]}$ and ${\mathfrak m} = 7.69\times 10^{10}~{\rm [Pa]}$. Here, we impose a low-reflecting boundary condition \cite{cohen1983silent} for letting waves pass out from the soil domain, which is defined as ${\bf \sigma}\cdot{\bf n} = -i\omega {\bf d}_{im}{\bf u}$
in the edges except for the incident perturbation edge in Fig. \ref{fig:2Dsimuls}A, B, and C. Here, ${\bf  n}$ is a unit normal vector, $\sigma$ is a stress tensor, and ${\bf d}_{im}$ is the mechanical impedance with ${\bf d}_{im} = {\bf d}_{im}(\rho, c_{s}, c_{p})$, where $c_{s}$ and $c_{p}$ are the speeds of shear waves and pressure in the material, respectively.

The simulation of the finite element method\footnote{This simulation was performed by using the {\tt COMSOL-MULTIPHYSICS 5.6}\textsuperscript{\texttrademark}.} for the frequency domain of $\{3.0, 6.5, 9.0\}$Hz with an incident displacement $u_{x}^{0}=0.3~[\rm m]$ shows that the incident wave from the displacement perturbation can be detoured only at $6.5$Hz of the bandgap frequency in the periodic structure, while other results demonstrate the opposite. To investigate the attenuation of the displacement amplitude, we set the cut line of the probe at $(x,0)$. Because the periodic structure is placed at $x~{\rm [m]}=(-11, 11)$, we may draw the root mean square (RMS) of the total displacement, $D_{\rm RMS}=\sqrt{\Delta x^2+\Delta y^2}$, along the cut line of the probe as shown in Fig. \ref{fig:2Dsimuls}D.
As a result, it is shown that the incident disturbance may therefore be successfully reduced in the metamaterial-like periodic structure region (Fig. \ref{fig:2Dsimuls}B and the green curve in Fig. \ref{fig:2Dsimuls}D).

\begin{figure*}[t!]  
\begin{center}
\includegraphics[width=\textwidth]{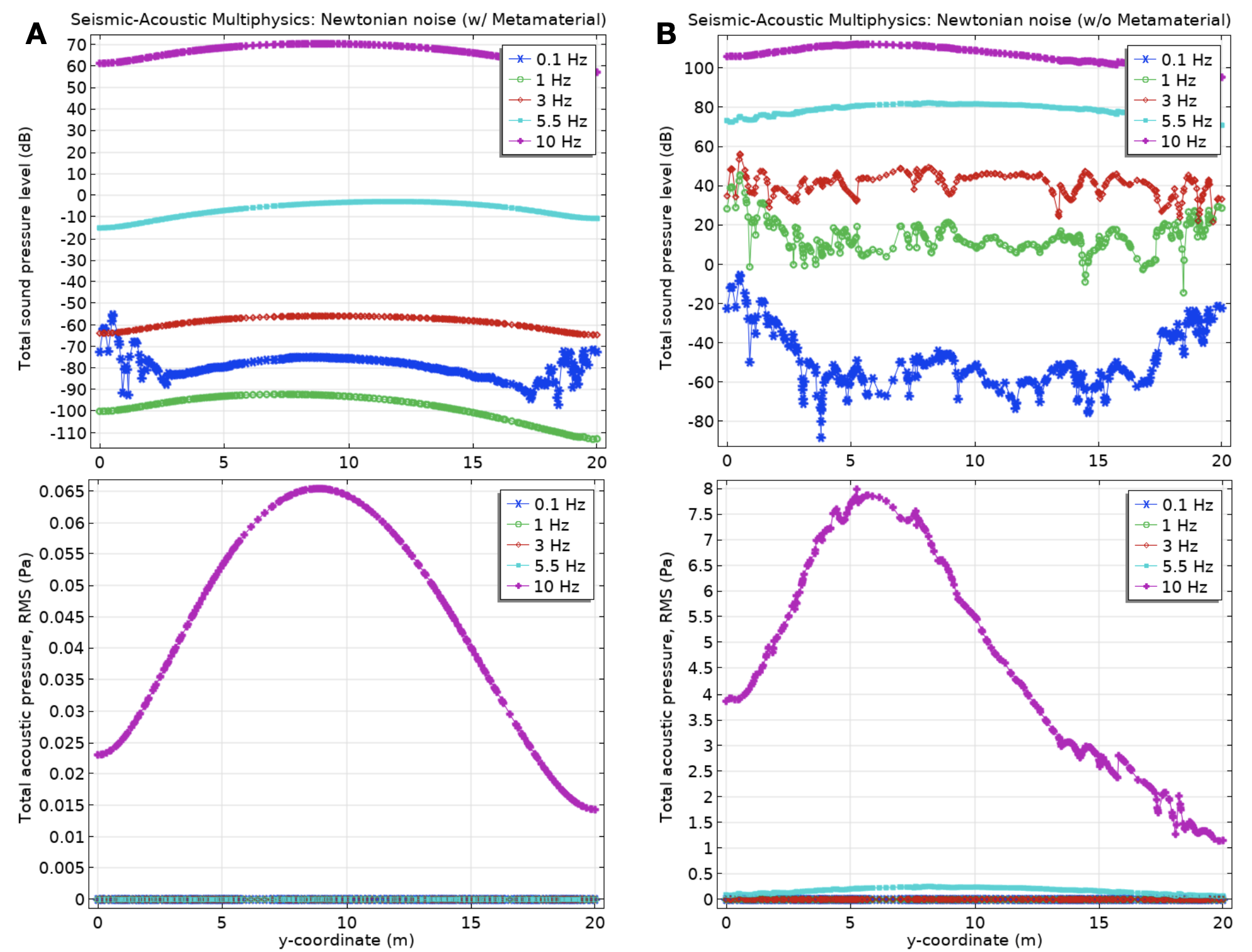}  
\caption{\small Plots of the total sound pressure level [dB] and the total acoustic pressure [Pa] along the probe cut line inside the tunnel with metamaterial-like beam strucure (A) and without metamaterial-like beam structure (B). For the badgap frequency of $\{0.1, 1.0, 3.0, 5.5\}$ Hz, the level of sound pressure and the acoustic pressure are effectively reduced. Moreover, there are still mitigated effects by the metamaterial-like structure for the $10$Hz case beyond the bandgap frequency range (violet curves).} \label{fig:KAGRA-MM2}
\end{center}
\end{figure*}



Next, we consider a three-dimensional design of the KAGRA \ac{gw} detector. For simplicity, we assume the $20$ m part of the full $3$ km arm length. A soil domain is established in which the cylinder beams were buried with the periodic structure of cylindrical beams near the underground tunnel of the KAGRA \ac{gw} detector. Similar periodic designs of cylinder beams have been considered and investigated in the Refs. \cite{doi:10.1142/9789813228702_0007,Miniaci_2016}. The basic design and setup we considered are illustrated in Fig. \ref{fig:KAGRA-MM1}A. In the soil domain, the $4$ m diameter underground tunnel with a concrete wall is considered, and there are periodic $0.3$ m radius structural steel cylinders close to the tunnel. 

The eigenvalue problem can be treated with a simple cell block shown in the LHS of Fig. \ref{fig:KAGRA-MM1}A, then the bandgap structure appears between $5.39\times 10^{-13} \rm Hz$ and $7.93 \rm Hz$ as shown in Fig. \ref{fig:KAGRA-MM1}B. Note that we impose the periodic boundary condition on all vertical surfaces and the low-reflecting boundary condition on the top/bottom surface of the cylinder.

Since we want to investigate the acoustic effect of seismic wave propagation, we need to solve the multiphysics analysis of structural and acoustic simulations. The solid mechanics for soil domain with structural steel beams and the concrete tunnel is governed by Eq. (\ref{eq:conteq}) with low-reflecting and free boundary conditions. 

On the other hand, the wave equation for sound waves in a lossless medium (no thermal conduction and no viscosity), in principle, is described by the inhomogeneous Helmholtz equation.
\begin{equation}
    \label{eq:sweq}
    \frac{1}{\rho c^2} \frac{\partial^2 p}{\partial t^2}+\nabla\cdot\left(-\frac{1}{\rho}(\nabla p -{\bf s}_{d})\right) = {s}_{m},
\end{equation}
where $p$ is the total pressure, $\rho$ is the total density, and $c$ is the speed of sound. Note that ${\bf s}_{d}$ and $s_{m}$ denote the sources of the dipole and monopole domains, respectively, which are set to zero in this study.

Indeed the vibration in the seismic domains of the soil, steel beam, and concrete tunnel with a metamaterial structure can affect the acoustic domain inside the tunnel. In the acoustic-structure boundary in the concrete tunnel, we have conditions,
\begin{eqnarray}
\label{acoustrucbc}
  && {\bf n}\cdot \left(-\frac{1}{\rho}(\nabla p - {\bf s}_{d})\right) = -{\bf n}\cdot \ddot{\bf u}, \\
  && {\bf F}_{A} = p{\bf n},
\end{eqnarray}
where ${\bf n}$ is a unit normal vector to the surface and ${\bf F}_{A}$ is the force projected on the area $A$.
Then the Helmholtz equation for sound waves combined with solid mechanical equation in this underground periodic structure can be solved in the frequency domain by a finite element approach with aforementioned boundary conditions. As a result, the computation with an incident displacement $u_{z}^{0}=1~[\rm m]$ has been done in the frequency values of $\{0.1, 1.0, 3.0, 5.5, 10.0\}$Hz and two simulation results of total sound pressure level inside the tunnel for $3$Hz and $10$Hz are shown in Figs. \ref{fig:KAGRA-MM1}C and \ref{fig:KAGRA-MM1}D. The total level of sound pressure within the tunnel for $3$Hz within the bandgap frequency is a negative value, while the total level of sound pressure for $10$ Hz outside the bandgap frequency range is quite loud inside the tunnel.

For a probe cut line at $(15, y, 5)$ as shown in Fig. \ref{fig:KAGRA-MM1}A, we draw the variation of the sound pressure level (top two figures of Fig. \ref{fig:KAGRA-MM2}) and the total acoustic pressure (RMS) (bottom two figures of Fig. \ref{fig:KAGRA-MM2}) through the tunnel inside. Here, the label A and B in Fig. \ref{fig:KAGRA-MM2} represents the case with a metamaterial-like beam and without the beam, respectively. Comparing two figures of Figs. \ref{fig:KAGRA-MM2}A and \ref{fig:KAGRA-MM2}B, it is found that the sound pressure level and the total acoustic pressure can be mitigated by the metamaterial-like structure. Furthermore, the mitigation appears to be effective even for the case of $10$Hz beyond the bandgap frequency range (the magenta curves in Fig. \ref{fig:KAGRA-MM2}).

In this study, we investigated a possibility of applying periodic metamaterial-like structures to the underground \ac{gw} detector by virtue of mitigating a low-frequency noise below $10$Hz. The suggested problem in this study was exhibited in the section III.C of \cite{Jung:2022kyc}; the non-linear correlation between the wind speed meter at the entrance of the KAGRA and the microphone channel in the physical environment monitor (PEM) system in the KAGRA detector during the daytime (Fig. 7 in \cite{Jung:2022kyc}). 

A possible assumption to explain this correlation is as follows; i) high wind between mountains during daytime strikes the slope of the mountain where the KAGRA is located, ii) the seismic vibration by wind strokes propagates to the underground site, iii) the seismic vibration transforms to the acoustic pressure perturbation at the concrete boundary of the KAGRA's tunnel, iv) the increased acoustic pressure perturbation produces the sound pressure noise inside the tunnel. 

The comparison between the $x$ and $y$ arms of the KAGRA detector for this assumption is shown in Fig. 8 of \cite{Jung:2022kyc}. Since the $y$-arm of the KAGRA detector is parallel to the slope of the mountain whereas the $x$-arm of the detector is stretched to deeper site from the slope, the presumed seismic vibration will be attenuated in the $x$-arm direction. Therefore, the acoustic noise converted from the seismic vibration by the winds becomes more severe in the $y$ arm tunnel, which may affect the \ac{gw} detection bands through the frequency up-conversion effect, which needs to be mitigated. Therefore, we investigated a possible way of using the metamaterial-like approach to mitigate seismic vibrations from the slope by inserting structured steel beams near the KAGRA tunnel. 

As a result, we have shown that the band gap structure of structural steel beams can reduce the appropriate sound pressure level compared to the case without the metamaterial-like structure. Moreover, it has been found that the mitigation effect appears even for the case beyond the bandgap frequency. Because the modeling and structure in this study are pretty simplified, further investigation with realistic and practical models and parameters should be required. 

Because the terrestrial \ac{gw} observatory is very sensitive to constant and transient seismic vibrations like an earthquake, this kind of low-frequency noise mitigation can be considered at the initial stage of construction. In addition, this approach is also worthwhile considering as a possible alternate way of the cancellation of Newtonian noises below $10$Hz (See Ref. \cite{Trozzo:2022tar} and references therein). Therefore, the relevant investigation for mitigating Newtonian noises and improving noise sensitivity at the low-frequency band may be required in future works. Furthermore, this study may potentially be applied to overcome the low-frequency noise barrier in the ground-based laser interferometer \ac{gw} telescope as well as the new conceptual ground-based \ac{gw} detector with the decihertz frequency bands like SOGRO \cite{Paik:2016aia,Paik:2019com}.









{\bf Acknowledgments}
The author thanks Piljong Jung, Edwin J. Son, Young-Min Kim, Chan Park, Muhan Choi, Dongwoo Lee, Hyung Won Lee, and Dong-Uk Hwang for helpful discussions and comments. This work was supported by the Basic Science Research Program through a National Research Foundation of Korea (NRF) funded by the Ministry of Education (NRF-2020R1I1A2054376) and the National Institute for Mathematical Sciences (NIMS).




\bibliographystyle{apsrev4-1}
\bibliography{Ref}
\end{document}